# The UV drag on hadronic hot jets as the origin of X-ray irradiation in AGN

## Karl Mannheim

Universitäts-Sternwarte, Geismarlandstr. 11, D – 37083 Göttingen, Germany
Internet: kmannhe@uni-sw.gwdg.de

**Abstract.** In this paper I calculate the anisotropic flux of cascade radiation produced by an isotropic distribution of highly relativistic protons immersed in the photon field of a spatially separated point source. Most radiation is produced towards the source of the target photons.

An astrophysical application of this result is X-ray irradiation of AGN accretion disks by highly dissipative nuclear jets. Low Mach number jets convert a large fraction of their kinetic power into relativistic particles via resonance with MHD turbulence. Consequently, these jets do not show up as radio jets on large scales.

Protons can easily reach high energies by stochastic acceleration in the turbulent jet plasma, whereas severe energy losses make it very difficult for the electrons to be accelerated at all. Proton acceleration saturates at the threshold energy for photo-production of secondary particles in the radiation field of the inner accretion disk. The energetic secondaries initiate an anisotropic electromagnetic cascade.

The backward cascade flux is reprocessed by the underlying accretion disk. The small fraction of cascade $\gamma$-rays emitted in the forward direction is consistent with $\gamma$-ray flux limits for Seyfert galaxies and with the spectrum of the diffuse cosmic photon flux at high energies.

## 1. Introduction

### 1.1. Relativistic protons and $\gamma$-rays

*In situ* observations of cosmic rays have shown that the energy density of highly relativistic particles in the interstellar medium is dominated by baryons and that the energy spectrum extends to energies of at least $10^{20}$ eV. This well-established empirical result has in the past led many authors (*e.g.*, Burbidge 1956) to claim that the relativistic particle populations in non-thermal extragalactic radio sources are also dominated by baryons.

It was soon recognized, however, that the leptons responsible for the radio synchrotron emission could not result as secondary particles from proton-matter collisions, since the matter density in extragalactic radio sources is by far too low (Perola 1969). It was therefore assumed that electrons can be efficiently accelerated as well as protons. The proton/electron energy density ratio depends on unknown details of the non-linear plasma processes which pre-accelerate particles above the threshold for the high energy acceleration mechanism.

However, the *coexistence* of highly relativistic protons and electrons in the plasma of extragalactic radio jets leads to a high energy extension of the radio/optical synchrotron emission due to X-/$\gamma$-ray synchrotron cascade emission (Mannheim *et al.* 1991). The low energy synchrotron photons constitute a thick target for the protons which cool by photoproduction of very energetic secondary particles.

Recently, CGRO and Whipple measurements have shown that radio jets oriented at small angles to the line of sight are indeed powerful sources of continuum spectra extending over almost 20 orders of magnitude in frequency – from radio frequencies to TeV photon energies (Fichtel *et al.* 1994, Punch *et al.* 1992). The energy flux of the $\gamma$-ray radiation can exceed the energy flux of the radio/optical synchrotron spectrum by a factor of the order of 100. Similarly, the energy density ratio of protons and electrons of the local cosmic ray particles is $u_p/u_e \simeq 100$. With such large values of the proton/electron ratio the proton-initiated cascade (PIC) model can explain spectra and temporal variability of the observed $\gamma$-ray emission (Mannheim & Biermann 1992, Mannheim 1993a,b).

At present, it is unclear whether jets at greater angles to the line of sight, i.e. steep spectrum radio sources, also show $\gamma$-ray emission as predicted. Experimentally, the problem is that the flux from jets pointing towards the observer is strongly amplified due to relativistic bulk motion of the ejected jet plasma and that therefore the $\gamma$-ray flux from jets at large angles is comparatively weak. At X-ray photon energies experiments are more sensitive and, indeed, Harris *et al.* (1994) succeeded in measuring X-ray emission from the hot spots in the jet of CygA with the ROSAT High Resolution Imager. The X-ray flux is in accord with the model prediction of proton-initiated cascades, but could also be due to synchrotron-self-Compton emission.

On the basis of these empirical facts it is proposed that the energy density of relativistic particles in jets is dominated by protons. It follows that jets must be considered as a source of high energy cascade emission contributing to the total emission of AGN.

### 1.2. AGN with disks and jets

Acceleration of nonthermal particles is most efficient in supersonic astrophysical jet flows. Jets could therefore hold the clue to the puzzling nonthermal properties of AGN. A key question is whether jets are universal in AGN as suggested by models where the angular momentum of the accreted material is transported away by a rotating MHD wind (Blandford & Payne 1982, Pudritz 1985, Lovelace *et al.* 1987, Königl 1989, Pelletier & Pudritz 1992, Ferreira & Pelletier 1993). If so, the nonthermal radiation of AGN could be radiation from an unresolved nuclear jet (Camenzind & Courvoisier 1983, Antonucci 1993, Siemiginowska & Elvis 1994, Mannheim 1994a). The high energy emission due to protons could then be very important.

In accord with the above models of coupled accretion/ejection systems the jets of radio-loud AGN not only seem to be energetically important, but they also seem to be energetically linked to the accretion flow feeding the putative supermassive black hole at the center. Rawlings & Saunders (1991) and Celotti & Fabian (1993) found the relation $L_{\rm nlr}/Q_{\rm j} \approx 0.01$ between the luminosity emitted in narrow forbidden lines $L_{\rm nlr}$ and the kinetic power of the jet $Q_{\rm j}$. The narrow lines re-emit only a very small fraction of the total luminosity of ionizing photons, i.e. typically $L_{\rm uv} = 100 L_{\rm nlr}$. Hence one obtains the striking relation

$$L_{\rm uv} \simeq Q_{\rm j}. \tag{1}$$

Remarkably, the emergence of powerful jets on large scales, which is characteristic for radio-loud AGN, cannot be read off the properties of the thermal emission from AGN of either type, radio-loud or radio-quiet. It is hard to imagine that the disk structure, presumably responsible for most of the thermal emission, does not bother the presence of a powerful jet at all. Moreover, it is obviously the case that the radiative losses of radio jets are very small compared to their kinetic power, for otherwise they would not be able to propagate over a distance of a Mpc[1]. This is almost miraculous and, as a matter of fact, it is quiet plausible to assume that the majority of jets suffers great energy losses by propagating through the central parsec of the host galaxy and never show up as a radio jet.

Indeed, if jets have a low Mach number, they become highly turbulent and dissipate much of their kinetic power radiatively (Bicknell & Melrose 1982, Pelletier & Zaninetti 1984). MHD turbulence transfers kinetic power via instabilities to resonant relativistic particles. One can therefore explain the absence of powerful radio jets in radio quiet AGN by an initially low Mach number of their jets (Mannheim 1994a). *Such turbulent jets would be hidden in the central parsec of the AGN.* Only a remnant biconical wind driven by the momentum transferred to turbulently entrained mass would survive on greater scales. Nuclear winds are common in AGN and their existence can solve the problem of the near-constancy of the ionization parameter from the BLR to the NLR (Smith 1993).

The first problem addressed in Sect.2 of this paper is to find out how the kinetic power of the putative nuclear jets in radio-quiet AGN is converted to high energy electromagnetic power. It is shown why protons might play such a crucial role. The jet cools by photoproduction of secondary particles in the UV photon field of the accretion disk. The second problem is to describe the spectrum of photons from the hadronic nuclear jet. Sect.3 treats the problem of the angular distribution of the proton-induced cascade radiation from the jet. Results apply to relativistic and non-relativistic jets and show the transition from forward to backward radiation with decreasing jet speed. Finally, Sect.4 discusses the results in the light of recent COMPTEL and EGRET limits of $\gamma$-ray fluxes from Seyfert galaxies and the question of the cosmic high energy background radiation.

## 2. Nuclear jets as proton accelerators

Due to MHD turbulence powerful low Mach number jets can dissipate away a large fraction of their kinetic energy by radiative cooling (Bicknell & Melrose 1982, Pelletier & Zaninetti 1984). Radio-quiet AGN could be equipped with low Mach number jets suffering shear and Kelvin-Helmholtz instabilities within the central parsec of their host galaxy. For such compact nuclear jets turbulent cascades transfer kinetic energy more likely to protons. *Due to their extreme energy losses electrons have difficulties to overcome the threshold energy for resonant interaction with MHD fluctuations – unless rapid processes like magnetic reconnection or interaction with Whistler waves pre-accelerate them efficiently.* While it is easy to think of a generalization of the model including electron acceleration, the emphasis of this paper is on the more likely proton acceleration. Note, however, that the basic result of

---

[1] Note that the comoving frame luminosity $L'_{\rm j} = D_{\rm j}^{-4} L_{\rm j}$ is much less than the apparent luminosity $L_{\rm j}$ inferred from

observations assuming isotropic unbeamed emission. $D_{\rm j} \lesssim 10$ denotes the Doppler boosting factor.

Sect.3, the anisotropy of proton induced high energy emission, is similarly obtained for Compton scattering electrons (Ghiselini *et al.* 1991).

The protons are accelerated up to the energy where photoproduction losses in the UV/soft X-ray field of the accretion disk set in. Photoproduced pions induce an electromagnetic cascade that irradiates the accretion disk from above. Another variant of this scenario would be a nuclear jet with focal shocks a few hundred Schwarzschild radii away from the central engine where shock acceleration operates (Pudritz 1992). In this physical picture jets may have to be considered as dissipative structures manifesting the outward transport of angular momentum of an accretion flow.

### 2.1. Physical conditions

Powerful jets establishing a centrifugal break on the accretion disk must be collimated by strong magnetic fields. The accretion disk radiates the luminosity

$$L_{uv} = \epsilon L_{Edd} = 1.3 \cdot 10^{45} \epsilon_{0.1} m_8 \text{ erg s}^{-1} \qquad (2)$$

where $\epsilon = 0.1\epsilon_{0.1}$ denotes an unknown efficiency factor of the accretion onto a black hole with mass $M = 10^8 m_8 M_\odot$. Equipartition between the energy densities of the thermal radiation field and the magnetic field at the base of the jet at $R_j \approx 10^2 r_2 R_G$ where $R_G = 1.5 \cdot 10^{13} m_8$ cm denotes the Schwarzschild radius yields the field strength

$$B = 2 \cdot 10^2 r_2^{-1} \epsilon_{0.1}^{\frac{1}{2}} m_8^{-\frac{1}{2}} \text{ G} \qquad (3).$$

If the jet survives passage through the central parsec of the host galaxy, it propagates to very large distances as a radio jet. The field strength measured at a scale of 1 kpc from Eq.(3) is $B(\text{kpc}) = 10^{-4}$ G, i.e. $B_{ds} = 4 \cdot 10^{-4}$ G in the downstream plasma of a strong shock with compression ratio $r = 4$ in agreement with radio and X-ray observations of hot spots (Meisenheimer *et al.* 1989, Harris *et al.* 1994).

Asserting that Eq.(1) holds true for AGN in general, i.e. that the thermal luminosity equals the kinetic power of a jet, one obtains for the density of protons in the jet

$$n_p = 3 \cdot 10^8 \epsilon_{0.1} m_8^{-1} \beta_{0.3}^{-3} r_2^{-2} \text{ cm}^{-3} \qquad (4)$$

where the jet speed is normalized to $\beta_j = 0.3\beta_{0.3}$. Eq.(4) entails particle conservation along the jet and holds true only for that part of the jet where mass entrainment can be neglected. For powerful radio jets mass conservation is probably valid over the entire length of the jet. The thermal matter density in the downstream plasma of Hot Spot shocks at the kpc scale is then given by $n_{p,ds}(\text{kpc}) = 3 \cdot 10^{-4}$ cm$^{-3}$ in agreement with observations.

For low Mach number jets $1 < M_j < 10$ shear, Kelvin-Helmholtz and kink instabilities establish a turbulent wake at a distance of $Z_{min} \gtrsim R_j M_j$ above the slow magnetosonic point, i.e. at

$$Z_{min} \gtrsim 100 M_j r_2 R_G. \qquad (5)$$

In contrast, high Mach number jets are protected from disrupting instabilities by a cocoon of hot backflowing material. Beyond $Z_{min}$ particles can scatter resonantly with the turbulence, if the particle momentum exceeds the threshold (Lacombe 1977)

$$p_\circ = m_p v_A c \simeq 80 \beta_{0.3}^{\frac{3}{2}} \text{ MeV/c} \qquad (6)$$

Acceleration of nonthermal particles is possible when the energy losses at $p_\circ$ are smaller than the energy gains due to scattering. The acceleration time scale is of the order of

$$t_{acc} \approx \frac{R_j}{v_A} \simeq 10^6 m_8 \beta_{0.3}^{-\frac{3}{2}} r_2 \text{ s} \qquad (7).$$

This time scale is relevant to 2$^{nd}$ order Fermi acceleration of particles interacting with fast magnetosonic waves generated by a Kraichnan cascade transferring energy released in Kelvin-Helmholtz or shear (or kink) instabilities towards dissipation scales. Including Alfvén-waves or other types of turbulence spectra introduces an energy dependence of the acceleration time scale and would therefore lead to a more complicated scenario. However, details of the resulting particle spectrum are irrelevant to the bottomline of the paper.

Now, as the jet lifts off the accretion disk it will diverge to some extent and the more tenuous the plasma becomes, the weaker are the energy losses at the threshold energy for acceleration. To obtain the minimum distance from the jet base where acceleration wins over energy losses one must compare $t_{acc}$ with the energy loss time scales for various cooling channels of the most relevant particle species, protons and electrons.

For protons energy losses due to Coulomb collisions with thermal electrons have to be considered. For electrons synchrotron and inverse-Compton losses are most important.

The proton threshold velocity $\beta_{p\circ}$ (in the comoving fluid frame) is nonrelativistic and is given by $\beta_p = 8.5 \cdot 10^{-2} \beta_{0.3}^{\frac{3}{2}}$. Due to turbulent and Compton heating the tenuous jet plasma is assumed to be very hot with a temperature of $T_j \approx 10^9$ K. This yields the Coulomb loss time for protons at the acceleration threshold (Mannheim & Schlickeiser 1994b)

$$t_{coul} = 1.3 \cdot 10^6 \epsilon_{0.1}^{-1} m_8 \beta_{0.3}^3 r_2^2 T_9^{\frac{3}{2}} \text{ s} \qquad (8)$$

Eq.(8) is valid as long as $T_9 = T_j/10^9$K $\gg 1.8 \cdot 10^{-2} \beta_{0.3}^3$. The bremsstrahlung luminosity of the thermal jet plasma is still unobservable, *viz.* $L_{ff} \ll L_j \lesssim L_{uv}$. For electrons the threshold momentum Eq.(6) is already relativistic and the synchrotron energy loss time scale is given by

$$t_{syn} = 42\epsilon_{0.1}^{-1} m_8 \beta_{0.3}^{-\frac{3}{2}} r_2^2 \text{ s} \qquad (9).$$

The energy distribution of the particles is given by the solution of a diffusion equation where the energy dependence of the diffusion coefficient is given by $D_\star(\gamma_p) = \gamma_p^2/(4t_{acc})$. Thus, Eqs.(8) and (9) have to be inserted into the condition for acceleration

$$4t_{acc} < t_{loss} \tag{10}$$

Eq.(10) is satisfied for protons accelerated at a jet radius

$$r_2 > 3\epsilon_{0.1}\beta_{0.3}^{-\frac{9}{2}}T_9^{-\frac{3}{2}} \tag{11}$$

and for electrons at

$$r_2 > 10^5\epsilon_{0.1} \tag{12}$$

Hence it follows that under the assumptions stated in the preceding paragraphs stochastic proton acceleration is possible from $Z = Z_{min}$ out to $Z = Z_{max}$ where the instabilities giving rise to the MHD turbulence have disrupted the jet flow. On the other hand, in the absence of a strong pre-acceleration mechanism electrons cannot be accelerated against the strong synchrotron (and inverse-Compton) drag. The efficiency of converting kinetic energy into protons (and, as shall be shown, into radiation) is given by

$$\chi = \frac{L_j}{Q_j} \approx M_j^{-3}\frac{Z_{max}}{R_j}\left[\gamma_{ad}\beta_P\right]^{-\frac{1}{2}} \tag{13}$$

where $\beta_P = p/2u_B \approx 1$ denotes the ratio of particle to magnetic pressure and $\gamma_{ad}$ is the adiabatic index of the plasma (Pelletier & Zaninetti 1984). From Eq.(13) it follows that the jet energy is dissipated within a range

$$10^2 R_G < Z_{max} < 10^5 R_G \tag{14}$$

for $1 < M_j < 10$ and $\beta_P = 1$. *Such nuclear jets are hidden in the central parsec of the AGN.*

Protons are accelerated up to the energy $\gamma_{p,max}$ where the condition $4t_{acc} = t_{cool,k}$ applies for the respective hadronic cooling channel $k$. In ascending order of the threshold energy $\gamma_{th,k}$ these channels are proton-proton pion production, photo-pair production and photo-pion production.

The cooling time scale for proton-proton pion production is given by

$$t_{pp} = [n\kappa_p\sigma_{pp}c]^{-1} = 7.4 \cdot 10^6\epsilon_{0.1}^{-1}m_8\beta_{0.3}^3r_2^2 \text{ s} \tag{15}$$

above

$$E_{th,pp} = 300 \text{ MeV} \tag{16}$$

where $\kappa_{pp} \simeq 0.5$ denotes the proton inelasticity and $\sigma_{pp} = 3 \cdot 10^{-26}$ cm$^2$ is the proton-proton hadronic cross section. The photo-pair production time scale is given by

$$t_{p\gamma,e^\pm} = \left[n_\gamma(> th,e^\pm)\kappa_{p,e^\pm}\sigma_{p\gamma,e^\pm}c\right]^{-1} = 3 \cdot 10^6\epsilon_{0.1}^{-1}m_8z_2^2x_{-4}^{-1} \text{ s} \tag{17}$$

where the inelasticity has the value $\kappa_{p,e^\pm} = 2m_e/m_p$ and the cross section reaches the value $\sigma_{p\gamma,e^\pm} = \frac{3}{8\pi}\alpha_f\sigma_T =$
6 · 10$^{-31}$ cm$^2$ at ten times the kinematic threshold for head-on collisions

$$\gamma_{p,th,e^\pm}^{(eff)} = 10 \cdot \frac{2}{(1-\mu_p)x_o} = 10^5 x_{-4}^{-1} \tag{18}$$

where $x_{-4} = x_o/10^{-4}$, $\mu_p = \cos\theta_p$ and $\theta_p$ is the scattering angle between photons and protons ($\mu_p = -1$ for head-on collisions). Finally, the photo-pion production time scale is given by

$$t_{p\gamma,\pi} = [n_\gamma(> th,\pi)\kappa_{p,\pi}\sigma_{p\gamma,\pi}c]^{-1} = 2 \cdot 10^4\epsilon_{0.1}^{-1}m_8z_2^2x_{-4}^{-1} \text{ s} \tag{19}$$

above the threshold energy for head-on collisions

$$\gamma_{p,th,\pi} = \frac{\frac{m_\pi}{m_e}(1+\frac{m_\pi}{2m_p})}{(1-\mu_p)x_o} = 1.5 \cdot 10^6 x_{-4}^{-1} \tag{20}$$

The inelasticity has the value $\kappa_{p,\pi} = 1/5$ and the cross section is $\sigma_{p\gamma,\pi} = 5 \cdot 10^{-28}$ cm$^2$ at threshold. The proton gyroradius at the energy Eq.(20) is given by

$$R_L = 2.3 \cdot 10^{10}\epsilon_{0.1}^{-\frac{1}{2}}m_8^{\frac{1}{2}}x_{-4}^{-1}r_2 \text{ cm} \ll R_j \tag{21}$$

and is thus consistent with the model assumptions. The photoproduction time scales are very short, but they become relevant only above the respective threshold energies. Therefore, condition Eq.(10) can be violated provided that $\gamma_p < \gamma_{p,th,k}$ and *acceleration saturates at the threshold energy*.

Comparing the energy loss time scales with the acceleration time scale one obtains the result that (i) proton-proton pion production does not limit the proton maximum energy for a jet radius

$$r_2 > 0.5\epsilon_{0.1}\beta_{0.3}^{-\frac{9}{2}} \tag{22}$$

The second result is that (ii) there is no photo-pair sphere inside of which the proton maximum energy would be given by Eq.(18), since $4t_{acc} > t_{e^\pm}$ holds only for

$$Z < Z_{min}M_j^{-1}\epsilon_{0.1}^{\frac{1}{2}}\beta_{0.3}^{-\frac{3}{4}}x_{-4}^{\frac{1}{2}}r_2^{-\frac{1}{2}} \tag{23}$$

However, there is a photo-pion sphere where acceleration saturates at the pion production threshold. It is defined by the range

$$Z_{min} < Z < Z_\pi = 1.4 \cdot 10^3\epsilon_{0.1}^{\frac{1}{2}}\beta_{0.3}^{-\frac{3}{4}}x_{-4}^{\frac{1}{2}}r_2^{-\frac{1}{2}}R_G \approx 14M_j^{-3}Z_{max}\epsilon_{0.1}^{\frac{1}{2}}\beta_{0.3}^{-\frac{3}{4}}x_{-4}^{\frac{1}{2}}r_2^{\frac{1}{2}} \tag{24}$$

where Eq.(13) has been used and the maximum energy is given by Eq.(20). Thus photo-pion production must be considered as the dominant radiative dissipation process under the conditions of a nuclear turbulent jet.

## 2.2. Cascade spectrum

Photo-pion production at the threshold energy has two channels of equal importance

$$p + \gamma \rightarrow \begin{cases} \pi^0 + p \rightarrow 2\gamma + p \\ \pi^+ + n \rightarrow e^+ + \nu_e + \nu_\mu + \bar{\nu}_\mu + n \end{cases} \quad (25)$$

Neutrons can be considered as stable over the scales of interest. The decay length is given by

$$Z_n = \gamma_n c t_n = (1 - \kappa_{p,\pi})\gamma_p c 10^3 \text{ s} \\ = 2.4 \cdot 10^6 m_8^{-1} x_{-4}^{-1} R_G \gg Z_{\max} \quad (26)$$

Neutrons which hit the disk initiate hadronic cascades bearing many similarities with extended air showers initiated by cosmic rays in the Earth's atmosphere. Due to the large penetration depth of relativistic particles their main effect is to heat the disk and to produce extra neutrinos which could be observable with underwater or underice detectors (Nellen *et al.* 1993).

High energy photons result from neutral pion decay, but also from synchrotron radiation and from inverse-Compton scattering of charged pion decay positrons. In the strong soft photon field from the accretion disk $\gamma$-rays are subject to photon-photon pair production

$$\gamma + \gamma \rightarrow e^+ + e^- \quad (27)$$

The cross section is maximized for

$$x_\gamma = \frac{E_\gamma}{m_e c^2} = \frac{2}{x_t} \quad (28)$$

where it has the value

$$\sigma_{\gamma\gamma} \simeq \frac{1}{3}\sigma_T \quad (29)$$

For $\gamma$-rays produced in the range

$$1 < x_\gamma < x_{\gamma,\max} = \frac{1}{2}\kappa_{p,\pi}\frac{m_p}{m_e}\gamma_{p,\max} = 2 \cdot 10^8 x_{-4}^{-1} \quad (30)$$

the relevant target photons are in the range

$$10^{-8} < x_t < 2 \quad (31)$$

corresponding to frequencies $10^{12-20}$ Hz. Therefore, in contrast to the target photons for photo-pion production which only come from the innermost part of the accretion disk, the target photons for $\gamma$-ray absorption come from the entire disk and from a possible dust torus. The photon distribution over this broad energy range shall be approximated by a power law

$$n(x_t) = n_0 x_t^{-2} = \frac{L_t}{4\pi Z^2 m_e c^3 \ln[x_{t,\max}/x_{t,\min}]} x_t^{-2} \quad (32).$$

This yields the photon-photon optical depth

$$\tau_{\gamma\gamma}(x_\gamma) = \frac{1}{3\Lambda_t} l x_\gamma \quad (33)$$

where $\Lambda_t = \ln[x_{t,\max}/x_{t,\min}]$ and the compactness is given by

$$l = \frac{\sigma_T}{4\pi m_e c^3}\frac{L_t}{Z} \\ = 0.1\epsilon_{0.1}^{\frac{1}{2}}\beta_{0.3}^{\frac{3}{4}}x_{-4}^{-\frac{1}{2}}r_2^{-\frac{1}{2}} \quad (34)$$

at $Z_\pi$ from Eq.(24). Inserting Eq.(34) into Eq.(33) gives

$$\tau_{\gamma\gamma}(x_\gamma) = 3 \cdot 10^{-3}\epsilon_{0.1}^{\frac{1}{2}}\beta_{0.3}^{\frac{3}{4}}x_{-4}^{-\frac{1}{2}}r_2^{-\frac{1}{2}}x_\gamma \quad (35)$$

where $\Lambda_t = 10$ has been taken. Thus high energy photons produced at the edge of the photo-pion sphere at $Z_\pi$ are absorbed for

$$x_\gamma \geq x_\gamma^* = 3 \cdot 10^3\epsilon_{0.1}^{-\frac{1}{2}}\beta_{0.3}^{-\frac{3}{4}}x_{-4}^{\frac{1}{2}}r_2^{\frac{1}{2}} \quad (35a)$$

and at the distance $Z_{\min}$ they are absorbed for

$$x_\gamma \geq x_\gamma^* = 10M_j\epsilon_{0.1}^{-1} \quad (35b).$$

Since the average photon energy of the decaying hadrons Eq.(30) is much higher, their electromagnetic power is reprocessed by photo-pair production to much lower energies. This yields a continuum flux density spectrum with spectral index close to unity, i.e. $\alpha_{\text{casc}} = 0.9 - 1.0$ up to the energy $x_\gamma^*$ where $\tau_{\gamma\gamma}(x_\gamma^*) = 1$ (Svensson 1987). According to Eqs.(35a) and (35b) the turnover energy $E_\gamma^* = x_\gamma^* m_e c^2$ should typically be found in the range from 10 MeV to 100 MeV.

## 2.3. Reprocessing by thermal matter

The fraction of cascade photons emitted towards the accretion disk at the base of the jet is further reprocessed by the thermal matter. (1) Hot electrons reflect part of the high energy irradiation, (2) photons above the Klein-Nishina turnover of the electron scattering cross section are thermalized and lead to significant heating of the disk surface, (3) iron ions re-emit fluorescent photons at $6.4 - 6.7$ keV and (4) the electrons of the plasma at the base of the jet constitute a non-static corona which comptonizes UV photons from the deeper and colder parts of the disk (Blandford & Payne 1981, Lightman & White 1988, Czerny & Elvis 1987, Pounds *et al.* 1990). The resulting spectra have been discussed in the literature extensively and no further calculations have to be presented here.

It shall only be noted that the proposition of ubiquitous jets in AGN solves the problem of the origin of comptonizing coronae. Soft X-ray excesses were found by ROSAT observations to be present much more common than previously assumed (Walter & Fink 1993, Walter *et al.* 1994). In the standard picture of $\alpha$-accretion disks this is only possible, if all the AGN accrete at nearly the Eddington rate. However, in the model presented here the inner part of the accretion disk where the UV photons originate is covered by the nuclear jet. Assuming $n_e = n_p$ one can readily calculate the electron scattering optical depth of the jet. For a ray with path length $R_j$ the optical depth is given by

$$\tau_e = n_e \sigma_T R_j \approx 0.3 \epsilon_{0.1} \beta_{0.3}^{-3} r_2^{-1} \tag{36}$$

The Compton $Y$-parameter is given by

$$Y = \frac{4kT}{m_e c^2} \mathrm{Max}[\tau_e, \tau_e^2] \approx 0.06 T_9 \epsilon_{0.1}^2 \beta_{0.3}^{-6} r_2^{-2} \tag{37}$$

Close to the accretion disk the the jet has a radius $r_2 \lesssim 0.1$ hence $\tau_e \lesssim 3$ and $Y \lesssim 2$. Such plasma conditions lead to a high energy wing of the UV-bump from the accretion disk by unsaturated comptonization of the soft photons with an X-ray flux density spectral index of $\alpha_x < 1$ (Rybicki & Lightman 1979). For those parts of the base of the jet where the particle bulk kinetic energy already exceeds the thermal energy the photon energy gain due to Comptonization goes at the expense of the bulk kinetic energy imposing a UV drag on the hot hadronic jet (Blandford & Payne 1981, Melia & König 1989, Siemiginowska & Elvis 1993).

## 3. The anisotropic proton initiated cascade

In this Section I calculate the energy and angular distribution of secondaries produced by a point source of high energy protons cooling in the monoenergetic photon field of a spatially separated point source of UV photons. This model demonstrates in a simple fashion the effects of (i) the relative velocity of the proton and photon sources $\beta_j$ and (ii) the difference k between the proton maximum energy and the head-on threshold energy for photo-production of pions on the anisotropy of the produced energy fluxes.

The assumptions are an approximation to the scenario that relativistic protons are generated at height $Z_{min} \geq M_j R_j \geq 100 M_j R_G$ in a jet lifting off perpendicular to an accretion disk and interacting with the photons from the inner edge of the disk at a radius of $R_i \approx 10 R_G$. In this case one obtains

$$\mu_o = \cos\theta_o = \left[ 1 + \left( \frac{R_i}{\gamma_j Z_{min}} \right)^2 \right]^{-\frac{1}{2}} \simeq 1 \tag{38}$$

for the cosine of the angle $\theta_o$ between photon momentum and the jet axis at the proton source location in the accretion disk frame. Moreover, in the comoving fluid frame of the jet, where quantities are denoted with a dash, one obtains

$$\mu_o' \simeq \mu_o \simeq 1 \tag{39}$$

For very large values of the jet Lorentz factor $\gamma_j$ even small deviations of $\mu_o$ from unity may become important to produce large differences to the value of $\mu_o'$. However, the main purpose of these calculations is to apply to the moderate relativistic regime. The approximation that only photons from the innermost ring of the disk or, equivalently, from the direction $\mu = \mu_o \simeq 1$ are counted at the jet location is justified by the fact that photoproduction of secondaries is dominated for reactions with target photons only above the threshold energy $x_o' \geq 280/(2\gamma_{p,max}')$. Thus, if acceleration saturates at the photoproduction threshold energy for head-on collisions with photons, which corresponds the lowest value of $\gamma_{p,max}'$, only the highest energy photons contribute to the cooling.

The Lorentz transformations between the disk and jet frames are

$$x = \gamma_j x' (1 + \beta_j \mu') \tag{40}$$

and

$$\mu = \frac{\mu' + \beta_j}{1 + \beta_j \mu'} \tag{41}$$

Let $q_\pi'(\gamma_\pi', \mu_\pi')$ denote the volume emissivity of pions in cgs units $\mathrm{cm}^{-3}\mathrm{s}^{-1}$ at dimensionless energy $\gamma_\pi' = E_\pi'/m_\pi c^2$ and at $\mu_\pi' = \cos\theta_\pi'$ where $\theta_\pi'$ denotes the angle of the pion emitted from the acceleration zone in the jet with respect to the jet axis (the normal to the disk) in the comoving fluid frame. Then the emissivity of pions is given by

$$q_\pi'(\gamma_\pi', \mu_\pi') = c \int_0^\infty dx' \int_1^\infty d\gamma_p' \int_{-1}^{+1} d\mu_p' \tag{42}$$
$$\cdot \left[ 1 - \beta_p' \mu_p' \right] n_\gamma'(x') n_p'(\gamma_p', \mu_p') \frac{d\sigma_{p\gamma,\pi}}{d\gamma_\pi' d\mu_\pi'}$$

From the results in Sect.2.1 it is clear that the proton Lorentz factor obeys $\gamma_p' \gg 1$ and hence $\beta_p' \simeq 1$, $n_\gamma'(x')$ denotes the monodirectional photon density with units $\mathrm{cm}^{-3}$ and, correspondingly, $n_p'(\gamma_p', \mu_p')$ the proton density and, finally, $d\sigma_{p\gamma,\pi}/d\gamma_\pi' d\mu_\pi'$ denotes the differential cross section for pion production. The differential cross section can be fairly approximated by

$$\frac{d\sigma_{p\gamma,\pi}}{dx_\pi d\mu_\pi} \simeq \sigma_{p\gamma,\pi} \delta\left[ \gamma_\pi' - \bar{\gamma}_\pi'(\gamma_p) \right] \delta\left[ \mu_\pi' - \mu_p' \right] \tag{43}$$

where $\sigma_{p\gamma,\pi} \approx 5 \cdot 10^{-28}$ cm$^2$ denotes the pion production cross section at threshold, $\bar{\gamma}_\pi' = \kappa_{p,\pi}' \frac{m_p}{m_\pi} \gamma_p' = a_\pi \gamma_p'$ with $a_\pi \approx 1.3$. The strong beaming of the produced secondaries and the isotropic decay of the $\Delta$-resonance in the proton rest frame justify the $\delta$-approximations in Eq.(43). Eq.(43) is limited to energies satisfying the threshold condition

$$\gamma_p' x' \left[ 1 - \beta_p' \mu_p' \right] \geq 280 \tag{44}$$

The proton energy distribution is assumed to be isotropic in the comoving fluid frame

$$n_p'(\gamma_p', \mu_p') = \frac{1}{2} n_{po}' (\gamma_p')^{-2} \tag{45}$$

The target photons in the accretion disk frame have the energy distribution

$$n_\gamma(x) = n_{\gamma o} \delta[x - x_o] \tag{46}$$

roughly approximating a thermal bump peaking at $x_\circ$.

The target photon distribution in the jet frame follows from Eqs.(39), (40) and (46)

$$n'_\gamma(x') = n'_{\gamma \circ} \delta \left[ \gamma_{\rm j} x'(1 + \beta_{\rm j}) - x_\circ \right] \qquad (47).$$

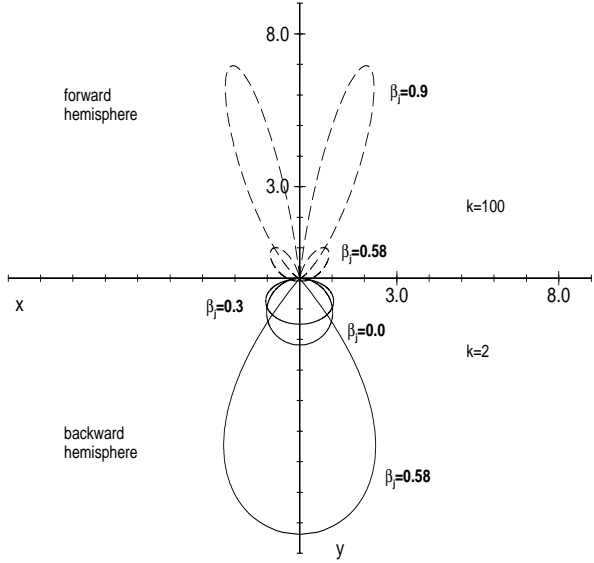

**Fig. 1.** Antenna diagram for photo-pion jet emission. The radius vector $r(\theta) = \sqrt{x^2 + y^2}$ gives the value of the normalized angular distribution of the pion power $\overline{Q_\pi(\mu_\pi)} = Q_\pi(\mu_\pi)/\int_{-1}^{+1} Q_\pi(\mu_\pi) d\mu_\pi$ parametrized by the jet velocity $\beta_{\rm j}$ and the maximum proton energy $k = \gamma_{\rm p,max}/\gamma_{\rm p,th,\pi}$. For $k = 2$ pion photoproduction is only possible into the backward hemisphere $\mu_\pi \leq 0$ (solid lines). Close to the critical jet velocity $\beta_{\rm c}$ the backward anisotropy becomes even stronger and developes a narrow lobe towards the source of the target photons at $\mu_\pi = -1$. In contrast, for $k = 100$ pion production is possible for almost all angles in the comoving frame. In this case a Doppler beaming pattern in the forward direction arises as $\beta_{\rm j} \to 1$ (dashed lines).

From here it is simple to proceed, because the integrations are each over a delta-distribution. Eq.(42) yields

$$q'_\pi(\gamma'_\pi, \mu'_\pi) \simeq \frac{c a_\pi}{2 \gamma_{\rm j}(1 + \beta_{\rm j})} n'_{\rm p \circ} n'_{\gamma \circ} \sigma_{\rm p \gamma, \pi} \left( \gamma'_\pi \right)^{-2} (1 - \mu'_\pi)$$
$$\cdot H \left[ \gamma'_\pi - \frac{280 a_\pi (1 + \beta_{\rm j}) \gamma_{\rm j}}{x_\circ (1 - \mu'_\pi)} \right] H \left[ a_\pi \gamma'_{\rm p,max} - \gamma'_\pi \right] \qquad (48)$$

so that the pion power at $\mu'_\pi$ is given by

$$Q'_\pi(\mu'_\pi) = m_\pi c^2 V'_{\rm p} \int_1^\infty q'_\pi \gamma'_\pi d\gamma'_\pi = \frac{m_\pi c^3 a_\pi}{2 \gamma_{\rm j}(1 + \beta_{\rm j})}$$
$$\cdot n'_{\rm p \circ} n'_{\gamma \circ} V'_{\rm p} \sigma_{\rm p \gamma, \pi} (1 - \mu'_\pi) \ln \left[ \frac{\gamma'_{\rm p,max} x_\circ (1 - \mu'_\pi)}{280(1 + \beta_{\rm j}) \gamma_{\rm j}} \right] \qquad (49)$$

where $V'_{\rm p}$ denotes the volume of the proton acceleration region. The received power in the accretion disk frame results from the transformation property

$$Q_\pi(\mu_\pi) = \frac{dE_\pi}{dt d\mu_\pi} = \frac{dE_\pi}{dE'_\pi} \frac{dt'}{dt} \frac{d\mu'_\pi}{d\mu_\pi} Q'_\pi(\mu'_\pi)$$
$$= D_{\rm j}^4(\mu_\pi) Q'_\pi(\mu'_\pi) \qquad (50)$$

with the Doppler factor

$$D_{\rm j}(\mu_\pi) = \frac{1}{\gamma_{\rm j}(1 - \beta_{\rm j} \mu_\pi)} \qquad (51).$$

The angle of the emitted pions as a function of its value in the accretion disk frame is given by

$$\mu'_\pi = \frac{\mu_\pi - \beta_{\rm j}}{1 - \mu_\pi \beta_{\rm j}} \qquad (52),$$

hence one obtains for the luminosity emitted into the the horizon bounded by $a \leq \mu_\pi \leq b$

$$L_\pi[a, b] = L_{\pi \circ} \int_a^b \left( 1 - \frac{\mu_\pi - \beta_{\rm j}}{1 - \mu_\pi \beta_{\rm j}} \right) D_{\rm j}(\mu_\pi)^4$$
$$\cdot \ln \left[ \frac{\gamma'_{\rm p,max} x_\circ \left( 1 - \frac{\mu_\pi - \beta_{\rm j}}{1 - \mu_\pi \beta_{\rm j}} \right)}{280(1 + \beta_{\rm j}) \gamma_{\rm j}} \right] d\mu_\pi \qquad (53).$$

The normalization $L_{\pi \circ}$ is fixed by the condition

$$L_\pi = L_\pi[-1, +1] \simeq L_{\rm j} \approx \chi L_{\rm uv} \qquad (54)$$

where $\chi$ denotes the efficiency of energy conversion, cf. Eq.(13). The logarithmic factor in Eqs.(49) and (53) expresses the threshold condition Eq.(43). Fig.1 shows the antenna diagram for the jet emission where the length of the radius vector gives the pion power at an angle $\theta_\pi = \arccos[\mu_\pi]$. At angles

$$\mu_\pi \geq 1 - \frac{2}{k} \left[ \gamma_{\rm j}(1 + \beta_{\rm j})(1 - \frac{2}{k} \beta_{\rm j} \gamma_{\rm j}) \right]^{-1} \qquad (55)$$

the threshold condition is not satisfied, hence no pions are produced in the forward hemisphere $\mu_\pi > 0$ for $k = \gamma_{\rm p,max}/\gamma_{\rm p,th,\pi} \leq 2$ and any $0 \leq \beta_{\rm j} \leq 1$. At a critical jet velocity $\beta_{\rm c}$ determined by

$$k \gamma_{\rm c}(1 + \beta_{\rm c}) \left[ 1 - \frac{2}{k} \beta_{\rm c} \gamma_{\rm c} \right] = 1 \qquad (56)$$

pion production ceases entirely, since no photons are available above threshold. For expample, at $k = 2$ the critical velocity is $\beta_{\rm c} = 0.6$ (cf. Fig.2). If, however, proton acceleration does not saturate at the threshold energy Eq.(20), i.e. $k \gg 1$, a Doppler boosting emission pattern arises for $\beta_{\rm j} \to 1$. For slow velocities $\beta_{\rm j} \to 0$ the maximum back-to-front ratio reached in the case of very high proton maximum energies has a value of

b : f $= L_\pi[-1, 0] : L_\pi[0, 1] \simeq 3 : 1$

which can be considered as a lower limit (Fig.2). As can be seen from Figs.1 and 2 photo-pion production exhibits a *natural anisotropy: pions are produced back towards the point source of target photons* when the proton maximum energy is just the threshold energy. Only when protons are accelerated at shocks far away from the central UV-photon source, the usual Doppler boosting emission pattern is generated by the jet. For the case of electrons an anisotropy similar to the $k \gg 1$ case of protons arises due to the kinematics of Compton scattering (Ghiselini et al. 1991).

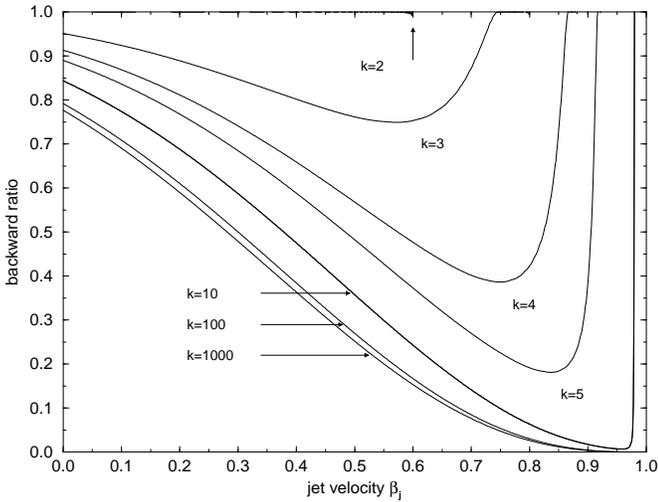

**Fig. 2.** The backward ratio $L_\pi[-1, 0] / L_\pi[-1, 1]$ vs. jet velocity $\beta_j$ parametrized by $k = \gamma_{p,max}/\gamma_{p,th,\pi}$ where $\gamma_{p,th,\pi}$ denotes the threshold energy for pion production in head-on collisions with photons. The luminosity peak is shifted to the hemisphere seen by the accretion disk as the jet speed decreases. In contrast, for relativistic speeds almost the entire secondary luminosity is emitted into the forward Doppler cone. At the critical speed $\beta_c$, however, the target photons as seen from the comoving fluid frame are redshifted out of the threshold condition Eq.(20) and the entire power goes backwards in the relativistic case also.

## 4. The high energy irradiation source

The presence of relativistic baryons inevitably causes high energy emission. AGN with relativistic jets pointing at small angles toward the terrestrial observer have indeed been shown to produce a large flux of $\gamma$-ray emission. The proposition of this paper is that all AGN have jets and that the radio-quiets have jets that become turbulent and dissipative in the central parsec. As a consequence they irradiate the accretion disk feeding the central black hole with back-to-front luminosity ratios in the range 75% to 100% for pure photo-pion production (Fig.2). A direct test of the model is to observe neutrino emission in the TeV-PeV range (Stecker et al. 1991). This is possible with planned underwater or underice detectors (Stenger et al. 1992). In contrast, high energy electromagnetic radiation is difficult to observe, because the large back-to-front ratios leave very little $\gamma$-ray flux in the line-of-sight.

Observational results from $\gamma$-ray satellites already impose strong limits on the presence of a spectral component with photon energies up to $10 - 100$ MeV in AGN (Maisack et al. 1993, Lin et al. 1993). Another constraint comes from the diffuse cosmic photon flux at high energies which could be due to the unresolved emission from AGN. In this Section I shall compare the model predictions with current experimental data.

### 4.1. Consistency with $\alpha_{ox}$ and $\gamma$-ray flux limits

The electromagnetic luminosity of the dissipative jet is given by

$$L_{j,em} = L_{\pi^0} + \frac{1}{4}L_{\pi^\pm} = \frac{5}{8}L_\pi \qquad (57)$$

where use has been made of $L_{\pi^\pm \to e^\pm} = \frac{1}{4}L_{\pi^\pm}$ (decay into four light particles, see Eq.(25)) and where $L_\pi$ denotes the total pion luminosity of the jet. The fraction of luminosity due to photoproduction of pairs is about

$$L_{e^\pm} \simeq \frac{t_{p\gamma,\pi}}{t_{p\gamma,e^\pm}} \ln\left[\frac{\gamma_{p,th,\pi}}{\gamma_{p,th,e^\pm}}\right] \simeq 0.02L_\pi \qquad (58)$$

using Eqs.(17)-(20) and is thus unimportant energetically. The same is true for the contribution of pions from proton-proton hadronic interactions derived from Eqs. (15),(16),(19) and (20)

$$L_{pp \to \pi} \simeq \frac{t_{p\gamma,\pi}}{t_{pp}} \ln\left[\frac{\gamma_{p,th,\pi}}{\gamma_{p,th,pp}}\right] \simeq 0.095L_\pi \qquad (59).$$

Hence it follows that the photo-pion luminosity $L_\pi$ equals the radiative power of the jet, i.e.

$$L_\pi \simeq L_j = \chi Q_j \qquad (60)$$

Some fraction of the kinetic power of the jet is not transferred to radiation, but is still in the kinetic energy reservoir of the turbulently entrained gas that ensures momentum conservation. The maximum efficiency is therefore given by $\chi \lesssim 1$ and one obtains from Eq.(1) that

$$L_{j,em} \lesssim \frac{5}{8}L_{uv} \qquad (61).$$

As shown in Sect.2.2 the proton-initiated cascade shifts the electromagnetic power into an energy range from the synchrotron-self-absorption energy

$$x_s = 1.4 \cdot 10^{-7} M_j^{-1} \epsilon^{\frac{2}{3}} r_2^{-\frac{4}{3}} m_8^{-\frac{1}{3}} \qquad (62)$$

(corresponding to $\nu_s = 2 \cdot 10^{13}$ Hz) up to the pair absorption energy $x_\gamma^\star$ derived from Eq.(35a). For the $\alpha_{casc} \simeq 1$ cascade flux density spectrum the value of the luminosity per logarithmic energy is given by

$$\frac{dL_{j,em}}{d\ln[x]} = \frac{L_{j,em}}{\ln[x_\gamma^\star/x_s]} = (0.04 - 0.05)L_{j,em} \lesssim 0.03L_{uv} \qquad (63).$$

*The nonthermal continuum from the dissipative jet is spread over a very large energy interval and is therefore apparently energetically unimportant in a particular narrow photon energy band.* Taking into account the back-to-front ratio of at least $3:1$ derived in Sect.3 one obtains for the forward energy flux

$$\frac{dL_{\rm j,em}}{d\ln[x]}({\rm forward}) \lesssim 0.008 L_{\rm uv} \qquad (64).$$

*Only the photons reprocessed by thermal matter in the vicinity of the jet re-radiate a significant fraction of $L_{\rm j,em}$ in the narrow X-ray band (Sect. 2.3).*

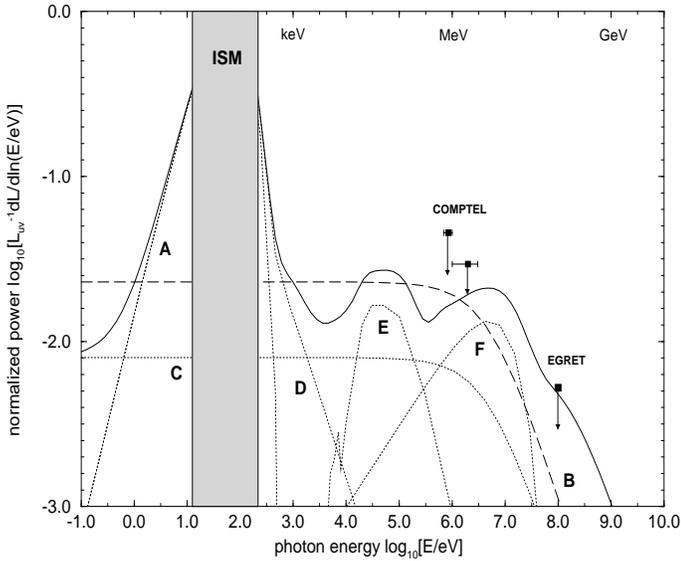

**Fig. 3.** A sketch of the generic AGN disk + nuclear jet broadband spectrum. The accretion disk is assumed to radiate the thermal spectrum marked **A**, the jet radiates spectrum **B** as seen from the accretion disk and **C** as seen in the forward direction (perpendicular to the disk). Spectra **D** and **E** arise from Comptonization of the UV photons at the base of the jet and from the reflection of jet radiation by the disk, respectively. Comptonization of disk photons by the hot plasma at the base of the nuclear jet provides for the major source of X-ray photons in the keV range and determines the value $\alpha_{\rm ox} \lesssim 1$. Finally, **F** denotes the synchrotron spectrum generated by the Bethe-Heitler photo-pairs in the jet. The **solid line** is the total spectrum as seen by an observer seeing the accretion disk face-on. For comparison the sketch also contains recent $\gamma$-ray flux limits for Seyfert galaxies normalized to the X-ray spectrum at 100 keV (Maisack *et al.* 1993, Lin *et al.* 1993).

Fig.3 shows a sketch of the standard multifrequency spectrum of a radio-quiet AGN as predicted by the model. In contrast to pure reflection models the X-ray emission in the 0.1-10 keV band is not direct nonthermal emission but it is a mixture of reflected and Comptonized emission from the base of the jet. The fraction of X-rays in the keV range resulting from Comptonization at the base of the jet

(D in Fig.3) to that of the irradiating X-rays (B in Fig.3) is of order unity explaining the commonly found value of the optical/X-ray spectral index $\alpha_{\rm ox} \lesssim 1$. The precise value depends upon details of the inclination angle and the temperature of the Comptonizing jet plasma.

The large values of the back-to-front ratio of the nonthermal emission from the jet are consistent with recent $\gamma$-ray flux limits for radio-quiet AGN. Since the disk also thermalizes a fraction of the order of the thermal disk luminosity itself from the nonthermal hadronic power emitted by the jet in the form of relativistic neutrons and $\gamma$-rays and since there is an underlying nonthermal power law reaching down to the optical range, simultaneous variability from the optical to the X-rays is expected (Clavel *et al.* 1992).

### 4.2. The diffuse high energy photon background

It has been proposed by various authors that the cosmic X-ray background is due to the unresolved emission from AGN. ROSAT measurements in the Lochman-hole seem to support this proposition (Hasinger 1992). X-ray reflection models can account for the high energy spectrum, when the back-to-front ratios are large enough to account for the steep turnover above 100 keV in the AGN rest frame due to the onset of the Klein-Nishina part of the cross section (Zdziarski *et al.* 1993) which is the case in the hadronic jet model proposed here for the origin of AGN X-rays. At higher energies a soft $\gamma$-ray bump at $\approx 5$ MeV[2] seems to be present in the diffuse background flux. The power ratio of the X-ray bump to the $\gamma$-ray bump is roughly (*e.g.*, Holt 1992)

$$\frac{dL_{\rm X}(30\,{\rm keV})}{d\ln[x]} \approx 2 \frac{dL_{\gamma}(5\,{\rm MeV})}{d\ln[x]},$$

which cannot be accounted for by the photo-pions (Fig.3). However, the photo-pairs produced with $\gamma_{\rm e,max} = \gamma_{\rm p,max}$ from Eq.(20) and with luminosity Eq.(58) indeed have the synchrotron $\gamma$-ray energy

$$x_{\gamma}({\rm e}^{\pm}) = 3.14 \cdot 10^{-14} B \gamma_{\rm e,max}^{2} \simeq 10 \epsilon_{0.1}^{\frac{1}{2}} m_8^{-\frac{1}{2}} x_{-4}^{-2} r_2^{-1} \qquad (65)$$

to be an interesting source of the $\gamma$-ray bump photons. The anisotropy described in Sect.3 is not relevant for these pairs, since they are produced far above threshold where target photons must be considered not only from the innermost ring of the accretion disk, but from greater radii as well. With the synchrotron spectrum of flux density

---

[2]     This is $\approx 17$ MeV in the AGN rest frame, if the Klein-Nishina interpretation of the hard X-ray turnover is correct.

index 1/2 the luminosity per logarithmic energy is given by

$$\frac{dL_\gamma[x_\gamma(e^\pm)]}{d\ln[x]} = 0.5L_\gamma \simeq \frac{4}{5}0.02L_{\mathrm{j,em}}$$
$$\simeq \frac{8}{15}\frac{dL_{\mathrm{j,em}}(\text{forward}, 100\text{keV})}{d\ln[x]} \quad (66)$$

obtained by using Eqs.(57),(58) and (63). This agrees with the observed value. To test the prediction $\gamma$-ray measurements of individual Seyfert galaxies in the 1-10 MeV range must be more sensitive by a factor of 2-5 than previously achieved.

## 5. Conclusions

A working hypothesis is presented in this paper: AGN represent accreting supermassive objects in which the angular momentum of the accreted gas is expelled through powerful MHD jets. Low Mach number jets cannot emerge out of the central parsec of their host galaxy. They rapidly develop MHD turbulence, entrain mass and slow down. The turbulence causes particle acceleration which leads to a high radiative efficiency for converting kinetic energy into radiation. I have argued that the dominant radiation mechanism of the nuclear dissipative jet could be photo-pion production by protons accelerated at a distance of $10^2 - 10^3 R_G$ above the disk. Photo-pion production in the photon field of the accretion disk displays a prominent anisotropy. Radiation is produced mostly in the direction of the target photons, i.e. back towards the inner accretion disk. As a consequence, the cold disk and the hot base of the jet act as a reprocessing medium for the high energy irradiation producing well-known features in the X-ray band such as Compton-reflection and Comptonization of UV-photons into the soft X-ray band. The shape of AGN X-ray spectra can in principal be explained by the model. Photo-pairs are energetically unimportant as a radiative energy loss of the jet. However, they are responsible for a weak soft $\gamma$-ray bump dominating the AGN spectrum in the 1-10 MeV range. The flux of the soft $\gamma$-ray bump is below the current CGRO flux limits for Seyfert galaxies. It is consistent with the diffuse $\gamma$-ray background emission, if the photo-pion induced X-ray emission reflected from AGN accretion disks is responsible for the diffuse X-ray background emission at around 30 keV.

Crucial experimental tests are the detection of $\gamma$-ray lines from the irradiated cold matter and the detection of high energy neutrinos which is feasible by experiments currently under construction.